\documentclass[12pt,a4]{article}
\usepackage{amsmath}

\usepackage{amssymb}
\begin{document}
\renewcommand{\theequation}{\thesection.\arabic{equation}}
\thispagestyle{empty}
\vspace*{-1.5cm}
\hfill {\small KL--TH 98/13} \\[8mm]

\setlength{\topmargin}{-1.5cm}
\setlength{\textheight}{22cm}
\begin{center}
{\large AN EXACTLY SOLVABLE MODEL OF THE \\

CALOGERO TYPE FOR THE ICOSAHEDRAL GROUP }\\

\vspace{2 cm}
{\large O. Haschke and W. R\"uhl}\\
Department of Physics, University of Kaiserslautern, P.O.Box 3049\\
67653 Kaiserslautern, Germany \\
\vspace{5cm}
\begin{abstract}
We construct a quantum mechanical model of the Calogero type 
for the icosahedral group as the structural group. Exact 
solvability is proved and the spectrum is derived explicitly.
\end{abstract}
\vspace{3cm}
{\it October 1998}
\end{center}
\newpage

\section{Introduction}
We have emphasized in \cite{1} that the exactly solvable models known so 
far \cite{2} are characterized by a structural group, which 
is a Coxeter group. The crystallographic property which turns a 
Coxeter group into a Weyl group \cite{3} can be abandoned. Doing 
this we neither have a simple Lie group at our disposal to analyse 
the model, nor the symmetric spaces etc. To verify our conjecture 
we consider the icosahedral group $H_3$ \cite{4} which is non-crystallographic.

The differential operator techniques to construct exactly solvable models are 
explained in detail in \cite{1}. In the weight space of 
$H_3$ which is a Euclidean space $\mathbb{R}^3$, we introduce Cartesian 
coordinates and construct from them the basis of 
invariants $\{I_n, \, n \in \{2,6,10\}\}$ (section 2). In section 3 we 
express the Laplacian in $\mathbb{R}^3$ in coordinates $\{I_n\}$. In general 
the determinant of the inverse Riemann tensor is a polynomial in these $\{I_n\}$ 
and factorizes in further real polynomials ("prepotentials") in the $\{I_n\}$, 
one factor for each orbit of the group $H_3$. However, $H_3$ has only one orbit. 
The factorization is therefore trivial. Thus we obtain one prepotential from 
the determinant, in the Calogero case we have in addition an oscillator 
prepotential. We return then to the most general Schr\"odinger operator. 
In section 4 we prove exact solvability of this operator and 
calculate its spectrum. A discussion of the problems arising in an attempt to 
define a Sutherland model can be found in section 5.
\setcounter{equation}{0}
\section{The invariants of $H_3$. }

The icosahedral group is generated by reflections $s_1, s_2, s_3$ denoting reflections along the simple weights
\begin{eqnarray}
\alpha_1 &=& (a, - \frac12, b) \label{2.1} \\
\alpha_2 &=& (-a, \frac12, b) \label{2.2} \\
\alpha_3 &=& (\frac12, b, -a) \label{2.3}
\end{eqnarray}
where
\begin{eqnarray}
a &=& \cos \frac{\pi}{5} = \frac14 (1 + \sqrt{5}) \label{2.4} \\
b &=& \cos \frac{2\pi}{5} = \frac14 (-1 + \sqrt{5}) \label{2.5}
\end{eqnarray}
A reflection $s_\alpha$ along a weight vector $\alpha$ acts on a vector $x$ according to
\begin{equation}
s_\alpha x = x - 2 \frac{(\alpha,x)}{(\alpha, \alpha)} \alpha \label{2.6}
\end{equation}
and leaves a hyperplane $H_\alpha$
\begin{equation}
(\alpha,x) = 0 \label{2.7}
\end{equation}
pointwise fixed.
$H_3$ consists of 60 rotations and 60 reflections which are generated from $s_1, s_2, s_3$ using the relations

\begin{equation}
s^2_1 = s^2_2 = s^2_3 = 1 \label{2.8}
\end{equation}
\begin{equation}
(s_2s_1)^5 = (s_3s_2)^3 = (s_3s_1)^2 = 1 \label{2.9}
\end{equation}
Among these 60 reflections are the reflections $s_4,s_5,s_6$ along the roots $\alpha_4, \alpha_5, \alpha_6$
\begin{eqnarray}
\alpha_4 &=& (1,0,0) \label{2.10} \\
\alpha_5 &=& (0,1,0) \label{2.11} \\
\alpha_6 &=& (0,0,1) \label{2.12} 
\end{eqnarray}
e.g.
\begin{equation}
s_4 = s_1s_2s_3s_2s_1 \label{2.13}
\end{equation}
Among the rotations of cyclicity 3 are the cyclic permutations
\begin{equation}
w = s_3s_2 = \left( \begin{array}{ccc}
0 & 0 & 1 \\
1 & 0 & 0 \\
0 & 1 & 0 \end{array} \right)
\label{2.14}
\end{equation}
\begin{equation}
w^{-1} = w^T = w^2 \label{2.15}
\end{equation}

Group actions on functions over the weight space are defined as usual
\begin{equation}
g \in H_3 : T_gf(x) = f(g^{-1}x) \label{2.16}
\end{equation}
so that for a polynomial
\begin{equation}
T_gp(x) = p(x), \, {\rm all} \, g \in H_3 \label{2.17}
\end{equation}
defines invariance. We can produce invariant polynomials by averaging over the group
\begin{equation}
p_{\rm inv}(x) = \frac{1}{120} \sum_{g \in H_3} T_gp(x) \label{2.18}
\end{equation}
This approach is simplified if the ansatz is already chosen invariant under 
$s_4,s_5,s_6,w$ and $w^2$. A possible (non unique) algebraic basis of 
invariant polynomials is
\begin{eqnarray}
I_2(x) &=& x^2_1 + x^2_2 + x^2 _3 \label{2.19} \\
\nonumber \\
I_6(x) &=& b(x^4_1x^2_2 + x^4_2x^2_3 + x^4_3x^2_1) \nonumber \\
& & - a (x^2_1 x^4_2 + x^2_2 x^4_3 + x^2_3 x^4_1) \nonumber \\
& & + 2 x^2_1 x^2_2 x^2_3 \label{2.20} \\
\nonumber \\
I_{10}(x) &=& -b(x^8_1x^2_2 + x^8_2x^2_3 + x^8_3x^2_1) \nonumber \\
& & - a(x^8_1x^2_3 + x^8_2x^2_1 + x^8_3x^2_2) \nonumber \\
& & - (b+2) (x^6_1x^4_3 + x^6_2x^4_1 + x^6_3x^4_2) \nonumber \\
& & + (2-a) (x^6_1x^4_2 + x^6_2x^4_3 + x^6_3x^4_1) \nonumber \\
& & + 6(a+b) (x^6_1x^2_2x^2_3 + x^6_2x^2_3x^2_1 + x^6_3x^2_1x^2_2) \nonumber \\
& & - 5(a+b) (x^4_1x^4_2x^2_3 + x^4_2x^4_3x^2_1 + x^4_3x^4_1x^2_2) \label{2.21}
\end{eqnarray}
If we combine the transposition
\begin{equation}
x_2 \leftrightarrow x_3 \label{2.22}
\end{equation}
with the sign change
\begin{equation}
\sqrt{5} \leftrightarrow - \sqrt{5} \label{2.23}
\end{equation}
then
\begin{equation}
I_{2,6} \leftrightarrow I_{2,6}, \; I_{10} \leftrightarrow - I_{10}. \label{2.24}
\end{equation}
At least a unique $I_ {10}$ can be defined by requiring this "symmetry".

\setcounter{equation}{0}
\section{The Riemannian}

We introduce the inverse Riemannian by
\begin{equation}
g^{-1}_{kl} = \sum^3_{i=1} \frac{\partial I_k}{\partial x_i} \frac{\partial I_l}{\partial x_i} \label{3.1}
\end{equation}
Each $g^{-1}_{kl}(x)$ is an invariant polynomial and can be expressed by the $\{I_n\}$ (Chevalley's theorem)
\begin{eqnarray}
g^{-1}_{2,n} &=& 2n I_n \label{3.2} \\
g^{-1}_{6,6} &=& - \sqrt{5} I_{10} - I_6I^2_2 \label{3.3} \\
g^{-1}_{6,10} &=& - I_{10}I^2_2 + 16 \sqrt{5} I^2_6I_2 - \sqrt{5}I_6I^4_2 \label{3.4} \\
g^{-1}_{10,10} &=& + 16 \sqrt{5} I_{10}I_6I_2 - \sqrt{5} I_{10}I^4_2 - 96 I^3_6 \nonumber \\
& & + 48 I^2_6 I^3_2 - I_6 I^6_2
\label{3.5}
\end{eqnarray}
The determinant is
\begin{eqnarray}
\det g^{-1} &=& 400 \sqrt{5} I^3_{10} - 400 I^2_{10} I_6I^2_2 + 16I^2_{10}I^5_2 \nonumber \\
& & + 5760 \sqrt{5} I_{10}I^3_6I_2 - 464 \sqrt{5} I_{10} I^2_6I^4_2 \nonumber \\
& & + 13824 I^5_6 - 11648 I^4_6 I^3_2 + 592 I^3_6I^6_2 \nonumber \\
& & - 16 I^2_6I^9_2
\label{3.6}
\end{eqnarray}
and does not factorize in real polynomials of $\{I_n\}$. But in original Cartesian 
coordinates it factorizes \cite{5}
\begin{eqnarray}
\det g^{-1} &=& 2^{28} \prod^3_{i=1} x^2_i  \prod_{\begin{array}{c} {\scriptstyle  \pm} \\ {\scriptstyle
(i,j,k) \, {\rm cycl.}} \\ {\scriptstyle {\rm permutations} \, {\rm of} (1,2,3)} \end{array}}
\left( \frac12 x_i \pm bx_j \pm ax_k \right)^2
\label{3.7} \\
\nonumber \\
&=& 2^{28} \prod_{{\rm pos.} \, {\rm weights} \, \alpha}
(\alpha,x)^2
\label{3.8}
\end{eqnarray}
We normalize the roots to length one.

As prepotentials we set
\begin{equation}
P_0 = e^{I_2} 
\label{3.9}
\end{equation}
\begin{equation}
P_1 = \det g^{-1}
\label{3.10}
\end{equation}
so that with
\begin{equation}
r^{(i)}_k = \sum_l g^{-1}_{kl} \frac{\partial}{\partial I_l} \log P_i
\label{3.11}
\end{equation}
we get
\begin{equation}
r^{(0)} = (2I_2, 6I_6, 10I_{10})
\label{3.12}
\end{equation}

\begin{equation}
r^{(1)} = (60, - 2I^2_2, 40 \sqrt{5} I_6 I_2 - 2\sqrt{5} I^4_2)
\label{3.13}
\end{equation}
The differential operator to start with is then
\begin{eqnarray}
D &=& - \sum_{k,l} \frac{\partial}{\partial I_k} g^{-1}_{kl} \frac{\partial}{\partial I_l} \nonumber \\
& & + \gamma_0 \sum_k r^{(0)}_k \frac{\partial}{\partial I_k} + \gamma_1 \sum_k r^{(1)}_k \frac{\partial}{\partial I_k}
\label{3.14}
\end{eqnarray}
with free coupling constants $\gamma_{0,1} \in \mathbb{R}$.

The gauge transformation
\begin{equation}
e^{-\chi} De^\chi = - \sum^3_{i=1} \frac{\partial^2}{\partial x^2_i} + W(x)
\label{3.15}
\end{equation}
is found from (see \cite{1})
\begin{equation}
\chi = \frac12 \sum_i (\gamma_i - \frac12) \log P_i(x)
\label{3.16}
\end{equation}
Since each factor $(\alpha,x)^2$ in (3.8) implies a second order pole $(\alpha,x)^{-2}$ in 
$W(x)$, $e^{-\chi}$ should possess a zero at $H_\alpha$ because the potential is repulsive. 
This implies due to (3.16)
\begin{equation}
\gamma_1 - \frac12 < 0
\label{3.17}
\end{equation}
The potential $W(x)$ is
\begin{eqnarray}
W(x) &=& (\gamma^2_0 - \frac14) \sum^3_{i=1} x^2_i \nonumber \\
& & + (\gamma^2_1 -  \frac14) \sum_{\mbox{all pos. weights \,$ \alpha$}} (\alpha,x)^{-2}
\label{3.18}
\end{eqnarray}

\setcounter{equation}{0}
\section{The polynomial eigenfunctions of $D$}

From (3.1) to (3.5) it seems to us a complicated guessing procedure to conceive a Lie algebraic 
approach to the eigenfunctions of $D$. We proceed rather by a mixed technique. 
First we separate a "radial" variable $R$ by
\begin{eqnarray}
R &=& I_2 \label{4.1}\\
S &=& I_6I_2^{-3} \label{4.2} \\
T &=& I_{10}I^{-5}_2
\label{4.3}
\end{eqnarray}
so that $D$ separates in a "radial" equation
\begin{eqnarray}
\left[ - 4 \frac{\partial}{\partial R} R \frac{\partial}{\partial R} + (60\gamma_1 - 32) \frac{\partial}{\partial R} + 2 \gamma_0R 
\frac{\partial}{\partial R} + \frac{\epsilon_{N,k}}{R} \right] \Phi(R) \nonumber \\
= E_{M,N,k} \Phi(R)
\label{4.4}
\end{eqnarray}
and an "angular" equation
\begin{equation}
\Delta(S,T) p_{N,k}(S,T) = \epsilon_{N,k} p_{N,k}(S,T)
\label{4.5}
\end{equation}
where
\begin{eqnarray}
\Delta(S,T) &=& \left( \frac{\partial}{\partial S}, \frac{\partial}{\partial T} \right) \Bigg( \begin{array}{l}
36 S^2 + S + \sqrt{5} T ,\\
- 16 \sqrt{5} S^2 + 60 ST + T + \sqrt{5} S,  \end{array} \nonumber \\
& & \begin{array}{l}
-16 \sqrt{5} S^2 + 60 ST + T + \sqrt{5} S \\
+ 96 S^3 + 100 T^2 - 48 S^2 - 16 \sqrt{5} ST + S + \sqrt{5} T \end{array} \Bigg)
\left( \begin{array}{c} \frac{\partial}{\partial S} \\
\frac{\partial}{\partial T} \end{array} \right) \nonumber \\
& & - \gamma_1 \left[ (180 S + 2) \frac{\partial}{\partial S} + (300 T - 40 \sqrt{5} S + 2 \sqrt{5}) \frac{\partial}{\partial T} \right]
\label{4.6}
\end{eqnarray}
The eigenfunctions of the Schr\"odinger operator (3.15) have the form
\begin{equation}
e^{-\chi} \Phi(R) P(S,T)
\label{4.7}
\end{equation}
We shall see at the end that the last two factors in (4.7) can be written as one polynomial in $\{I_n\}$.

First we analyse the angular equation (4.5). We define a space of real polynomials $V_N(3N \in \mathbb{Z}_{\ge})$
\begin{eqnarray}
V_N &=& {\rm span} \big\{ S^{n_1} T^{n_1}; {\rm deg} (S^{n_1}T^{n_2}) \nonumber \\
&=& \frac23 n_1 + n_2 \le N \Big\}
\label{4.8}
\end{eqnarray}
We call a polynomial in $S,T$ "homogeneous of degree r" if all of its monomials have
\begin{equation}
{\rm deg} (S^{n_1} T^{n_2}) = r
\label{4.9}
\end{equation}
Then $\Delta$ can be expanded (finitely)
\begin{equation}
\Delta = \Delta^{(0)} + \Delta^{(1)} + \Delta^{(2)} + ...
\label{4.10}
\end{equation}
so that
\begin{equation}
{\rm deg} (\Delta^{(r)} p_m) = m - \frac r3
\label{4.11}
\end{equation}
if $p_m$ is homogeneous of degree $m$. A term such as
\begin{equation}
\frac{\partial}{\partial T} ST \frac{\partial}{\partial T}
\label{4.12}
\end{equation}
is typical for $\Delta^{(1)}$. We diagonalize $\Delta^{(0)}$ on $V_N/V_{N-\frac13}$ first.

Let $\xi \in V_N/V_{N-\frac13}$ be expanded
\begin{equation}
\xi = \sum^{\left[ \frac{N}{2}-q \right]}_{n=0} \alpha_n T^{N-2(n+q)} S^{3(n+q)}
\label{4.13}
\end{equation}
where
\begin{equation}
N \hat{=} 2 q \, {\rm mod} \, 1
\label{4.14}
\end{equation}
The eigenvalue equation
\begin{equation}
\Delta^{(0)}\xi = \epsilon \xi
\label{4.15}
\end{equation}
leads to the recursion relation
\begin{eqnarray}
4 (5N-n-q)(5N-n-q+8-15 \gamma_1) \alpha_n \nonumber \\
+ 96 (N - 2(n+q-1))(N - 2n - 2q + 1) \alpha_{n-1} = \epsilon \alpha_n
\label{4.16}
\end{eqnarray}
The eigenvalues are therefore
\begin{eqnarray}
\epsilon_{N,k} = 4(5N-k-q)(5N-k-q+8-15 \gamma_1) \nonumber \\
0 \le k \le \left[ \frac {N}{2} -q \right]
\label{4.17}
\end{eqnarray}
Due to (3.17) the eigenvalues are positive and (for fixed N only) nondegenerate. 
The corresponding eigenvector $\xi_{N,k}$ has expansion coefficients
\begin{eqnarray}
\alpha_n &=& 0, n < k \nonumber \\
\alpha_k &=& 1 \nonumber \\
\alpha_n &=& \prod^n_{\begin{array}{c} {\scriptstyle
m=k+1} \\ {\scriptstyle (n > k)} \end{array}}
\frac{96(N-2(m+q)+2)(N-2(m+q)+1)}{\epsilon_{N,k} - \epsilon_{N,m}}
\label{4.18}
\end{eqnarray}

The full eigenfunction $p_{N,k} \in V_N$ is split in two pieces
\begin{equation}
p_{N,k} = \xi_{N,k} + \tilde{p}_{N,k}
\label{4.19}
\end{equation}
\[ \xi_{N,k} \in V_N/V_{N-\frac13}, \; \tilde{p}_{N,k} \in V_{N-\frac13} \]
Similarly we decompose $\Delta$
\begin{equation}
\Delta = \Delta^{(0)} + \tilde{\Delta}
\label{4.20}
\end{equation}
Then it follows
\begin{equation}
\tilde{p}_{N,k} = \sum^{n_0(N)}_{n=1} \left[ (\epsilon_{N,k} - \Delta^{(0)})^{-1} \tilde{\Delta} \right]^n \xi_{N,k}
\label{4.21}
\end{equation}
Since each power of
\[ (\epsilon_{N,k} - \Delta^{(0)})^{-1} \tilde{\Delta} \]
lowers the degree by $\frac13$ at least, we have
\begin{equation}
n_0(N) = 3N
\label{4.22}
\end{equation}
However, one or more terms in (4.21) may be infinite if
\begin{equation}
\epsilon_{N^{\prime},k^{\prime}} = \epsilon_{N,k} \; \mbox{for some} \; N^{\prime} < N
\label{4.23}
\end{equation}
Then the eigenvector of the larger $N$ must be skipped. In general, from a sequence of degenerate eigenvalues
\begin{eqnarray}
\epsilon_{N,k} = \epsilon_{N^\prime, k^\prime} = \epsilon_{N^{\prime\prime},k^{\prime\prime}} = ... \nonumber \\
N < N^\prime < N^{\prime\prime} < ...
\label{4.24}
\end{eqnarray}
only the lowest $N$ contributes an eigenfunction. As a consequence, the Schr\"odinger operator (3.15), (3.16) is selfadjont in a space smaller than an $L^2$-space.

Finally we turn our attention to the radial equation (4.6). The index equation is
\begin{equation}
- 4 \alpha^2 + (60 \gamma_1-32) \alpha + \epsilon_{N,k} = 0
\label{4.25}
\end{equation}
implying
\begin{equation}
2 \alpha_\pm = (15\gamma_1 - 8) \pm [(15 \gamma_1-8)^2 + \epsilon_{N,k}]^{\frac12}
\label{4.26}
\end{equation}
Inserting (4.17) into (4.26) we obtain
\begin{equation}
\alpha_\pm =  
\begin{cases}  5N-k-q \\
(15\gamma_1-8) - (5N-k-q)
\end{cases}
\label{4.27}
\end{equation}
Regular solutions correspond to $\alpha_+$. It follows
\begin{equation}
\Phi(R) = R^{5N-k-q} \,  _1F_1(5N-k-q - \frac{E}{2\gamma_0}, 9-15\gamma_1 + 2(5N-k-q); \; \frac12 \gamma_0R)
\label{4.28}
\end{equation}
Due to the factor
\begin{equation}
e^{-\chi} = e^{- \frac12 (\gamma_0- \frac12)R} \times \mbox{ polynomial factor}
\label{4.29}
\end{equation}
and the asymptotic behaviour
\begin{equation}
_1F_1 \tilde{=} \, e^{\frac12 \gamma_0R} \times \mbox{ polynomial factor}
\label{4.30}
\end{equation}
the hypergeometric series must terminate
\begin{eqnarray}
E = E_{M,N,k} = 2 \gamma_0 (M + 5N - k-q) \nonumber \\
(M \in \mathbb{Z}_{\ge})
\label{4.31}
\end{eqnarray}
Thus the regular eigenfunctions of $D$ are polynomials in $R,T,S$ which inserting 
(4.1) - (4.3) turn into polynomials in $I_2, I_6, I_{10}$.

\setcounter{equation}{0}
\section{Problems with an $H_3$ Sutherland model}

We define trigonometric invariants for $H_3$ by sums over invariant subsets of the root set $R$ 
(remember the normalization of all roots is $(\alpha,\alpha) = \parallel \alpha \parallel^2 = 1$)
\begin{equation}
F_1(x) = \frac{1}{30} \sum_{\alpha \in R} e^{i(\alpha,x)}, \quad ( \#R = 30)
\label{5.1}
\end{equation}
%
%
\begin{equation}
F_{2,\xi}(x) = \frac{1}{N_\xi} \sum_{\{\alpha,\beta\} \in R \times R} \delta_{(\alpha,\beta), \xi} e^{i(\alpha + \beta,x)}
\label{5.2}
\end{equation}
\begin{eqnarray}
F_{3,\xi_1\xi_2\xi_3}(x) = \frac{1}{N_{\xi_1\xi_2\xi_3}} \sum_{\{\alpha,\beta\,\gamma\} \in R \times R \times R} \delta_{(\alpha,\beta), \xi_1} \delta_{(\beta, \gamma),\xi_2} \delta_{(\gamma,\alpha),\xi_3}  e^{i(\alpha + \beta + \gamma,x)} 
\label{5.3}
\end{eqnarray}
etc. We leave it unproven whether $n$-tupels of roots with fixed angles between them 
constitute one or more than one orbit. All invariants are normalized so that
\begin{equation}
F_1(0) = F_{2,\xi}(0) = F_{3,\xi_1\xi_2\xi_3}(0) = ... = 1
\label{5.4}
\end{equation}
Instead of cosines of angles we can characterize such $n$-tupels by lengths
\begin{eqnarray}
\parallel \alpha + \beta \parallel^2 &=& 2 (1+\xi) \label{5.5} \\
\parallel \alpha + \beta + \gamma \parallel^2 &=& 3 + 2 (\xi_1+\xi_2+ \xi_3) \label{5.6}
\end{eqnarray}
The angles assume only finitely many values
\begin{equation}
\xi \in \{ \pm 1, \pm \frac12, \pm a, \pm b, 0 \}
\label{5.7}
\end{equation}
where $\xi = -1$ can be eliminiated because it is trivial.

Denote, say
\begin{eqnarray}
F_1(x) &=& 1 + \phi_1(x) \label{5.8} \\
F_{2,0}(x) &=& 1 + \phi_2(x) \label{5.9} \\
F_{3,000}(x) &=& 1 + \phi_3(x) \label{5.10}
\end{eqnarray}
then the $\phi_i$ can be expanded into a power series in the polynomial invariants $I_n(x), \, n \in \{2,6,10\}$
\begin{equation}
\phi_i = \sum_{\begin{array}{c} {\scriptstyle
k,l,m \in \mathbb{Z}_\ge} \\ {\scriptstyle k+l+m > 0} \end{array}}
d^{(i)}_{klm} \frac{I^k_2I^l_6I^m_{10}}{(2k+6l+10m)!}
\label{5.11}
\end{equation}
These series are entire analytic and invertible in a neighborhood of zero. The inverse 
functions have branch cuts. In the Appendix we give the coefficients $d^{(i)}_{klm}$ for 
$2k + 6l + 10m \le 20$.

The invariants
\[ \phi_1, \phi_2, \phi_3 \]
can be used as coordinates for the Riemann tensor. From
\begin{eqnarray}
G_{kl}^{-1} = \sum_{i=1}^{3} \frac{\partial{\phi_k}}{\partial{x_i}} \frac{\partial{\phi_l}}{\partial{x_i}} \label{5.11a}
\end{eqnarray}
we obtain e.g.
\begin{equation}
G^{-1}_{11} = -\frac{1}{(\#R)^2} \sum_\xi \xi N_\xi F_{2,\xi}
\label{5.12}
\end{equation}
\begin{equation}
G^{-1}_{12} = -\frac{1}{\#R \cdot N_0} \sum_{\xi_1,\xi_3} (\xi_1 + \xi_3) N_{\xi_1,0,\xi_3} F_{3,\xi,0,\xi_3}
\label{5.13}
\end{equation}
etc. Between $F_1$ and the $F_{2,\xi}$ there are two relations
\begin{equation}
F_{2, - \frac12} = F_1
\label{5.14}
\end{equation}
\begin{equation}
F^2_1 = \sum_\xi  \frac{N_\xi}{(\#R)^2} F_{2,\xi}
\label{5.15}
\end{equation}
Since (5.12) involves
\begin{equation}
F_{2,a} - F_{2,-a}, F_{2,b} - F_{2,-b}
\label{5.16}
\end{equation}
and in (5.15) we recognize the combinations
\begin{equation}
F_{2,a} + F_{2,-a}, F_{2,b} + F_{2,-b}
\label{5.17}
\end{equation}
there are not enough polynomial relations to express $G^{-1}_{11}$ as a polynomial in the 
$\{ \phi_i \}$. The expansion is infinite.
%
%
%

%
%
%
\setcounter{equation}{0}
%
\section{Appendix}

In order to enable the reader to test the result of section 5 we give the expansion coefficients of 
$F_1,F_{2,0},F_{3,0,0,0}$ (\ref{5.1})--(\ref{5.3}), and $G_{11}^{-1}$ to the 
order 20.
\begin{eqnarray}
G_{11}^{-1}= \sum_{\begin{array}{c} {\scriptstyle
k,l,m \in \mathbb{Z}_\ge} \\ {\scriptstyle k+l+m > 0} \end{array}}
e^{(11)}_{klm} \frac{I^k_2I^l_6I^m_{10}}{(2k+6l+10m)!} \label{6.1}
\end{eqnarray} 
\begin{center}
\begin{tabular}{|c|c|c|c|c|} \hline
$k,l,m$ & $ d^{(1)}$ & $d^{(2)}$ & $d^{(3)}$ & $e^{(11)}$ \\ \hline
$1,0,0$ & $\frac{-1}{3}$ & $\frac {-2}{3}$ & -1 & 6 \\ \hline
$2,0,0$ & $\frac {1}{5}$ & $\frac {4}{5}$ & $\frac {9}{5}$ & $\frac {-72}{5}$ \\ \hline 
$0,1,0$ & $\frac {-1}{8}$ & $\frac {13}{8}$ & -6 & $\frac {27}{2}$ \\ \hline  
$3,0,0$ & $\frac {-7}{48}$ & $\frac {-53}{48}$ & -4 & $\frac {747}{20}$ \\ \hline 
$1,1,0$ & $\frac {7}{30}$ & $\frac {-91}{15}$ &$ \frac {168}{5}$ & $\frac {-546}{5}$ \\ \hline  
$4,0,0$ & $\frac {7}{60}$ & $\frac {49}{30}$ & $\frac {49}{5}$ & -105 \\ \hline 
$0,0,1$ & $ \frac {-5}{192} \sqrt{5}$ & $ \frac {5}{192} \sqrt{5}$ & $10 \sqrt{5}$ & $ \frac {111}{64} \sqrt{5}$ \\ \hline 
$2,1,0$ & $\frac {-49}{192}$  & $\frac {3073}{192}$ & -154 & $\frac {189903}{320}$ \\ \hline
$5,0,0$ & $ \frac {-19}{192} $ & $\frac {-485}{192}$  & -25 & $\frac {100389}{320}$ \\ \hline
$0,2,0$ & $\frac {261}{1280}$ & $\frac {-6111}{1280}$  & $\frac {-36}{5}$ &$ \frac {-56943}{160}$ \\ \hline
$1,0,1$ & $\frac {107}{1280} \sqrt{5}$ &$ -\frac {457}{1280} \sqrt{5}$ & $-\frac {432}{5} \sqrt{5}$ &$ -\frac {8631}{160} \sqrt{5}$ \\ \hline
$3,1,0 $& $\frac {1}{5}$ & $ \frac {-5887}{160}$ &$ \frac {3204}{5}$ & $\frac {-216207}{80}$ \\ \hline
$6,0,0 $& $\frac {449}{5120}$ & $\frac {20621}{5120}$ & $\frac {324
}{5}$ & $\frac {-628299}{640} $\\ \hline
$1, 2, 0 $&$ \frac {-2639}{3840}  $&$
 \frac {61789}{1920}  $&$\frac {364}{5}  $&$ \frac {2012283}{400} $\\ \hline
$2, 0, 1 $&$ -  \frac {637}{3840} \sqrt{5} $&$ \frac {3731}{1920} \sqrt{5} $&$ \frac {2366}{5} \sqrt{5} $&$ \frac {31941}{50} \sqrt{5} $\\ \hline
$4, 1, 0 $&$ \frac {-91}{960}  $&$ \frac {1183}{15}  $&$ \frac {-12376}{5}  $&$ \frac {4420143}{400} $\\ \hline
$7, 0, 0 $&$ \frac {-247}{3072}  $&$ \frac {-10075}{1536}  $&$-169 $&$ \frac {203931}{64} $\\ 
\end{tabular} \\
\end{center}
\begin{center}
\begin{tabular}{|c|c|c|c|c|} 
$0, 1, 1 $&$ -  \frac {29}{3840} \sqrt{5} $&$ -  \frac {8291}{3840} \sqrt{5} $&$ - \frac {1152}{5} \sqrt{5} $&$ \frac {302091}{256} \sqrt{5} $\\ \hline
$2, 2, 0 $&$ \frac {369}{256}  $&$ \frac {-32661}{256}  $&$144 $&$ \frac {-52597647}{1280} $\\ \hline
$3, 0, 1 $&$ \frac {2023}{7680} \sqrt{5} $&$ - \frac {56623}{7680} \sqrt{5} $&$ - \frac {10528}{5} \sqrt{5} $&$ - \frac {13333971}{2560} \sqrt{5} $\\ \hline
$5, 1, 0 $&$ \frac {-283}{7680}  $&$ \frac {-1247717}{7680}  $&$ \frac {45088}{5}  $&$ \frac {-106849329}{2560} $\\ \hline
$8, 0, 0 $&$ \frac {581}{7680}  $&$ \frac {83269}{7680}  $&$ \frac {2209}{5}  $&$ \frac {-27281247}{2560} $\\ \hline
$0, 3, 0 $&$ \frac {-3281}{40960}  $&$ \frac {-2492863}{40960}  $&$ \frac {9384}{5}  $&$ \frac {1119815109}{51200} $\\ \hline
$1, 1, 1 $&$ \frac {1071}{40960} \sqrt{5} $&$ \frac {559521}{40960} \sqrt{5} $&$ \frac {15912}{5} \sqrt{5} $&$ \frac {1038235509}{51200} \sqrt{5} $\\ \hline
$3, 2, 0 $&$ \frac {-198033}{81920}  $&$ \frac {32342721}{81920}  $&$ \frac {-28764}{5}  $&$ \frac {26545100517}{102400} $\\ \hline 
$4, 0, 1 $&$ -  \frac {30141}{81920} \sqrt{5} $&$ \frac {1886541}{81920} \sqrt{5} $&$ \frac {41616}{5} \sqrt{5} $&$ \frac {3555444753}{102400} \sqrt{5} $\\ \hline
$6, 1, 0 $&$ \frac {29223}{163840}  $&$ \frac {53514249}{163840}  $&$ \frac {-156978}{5}  $&$ \frac {30155348493}{204800} $\\ \hline
$9, 0, 0 $&$ \frac {-14263}{196608}  $&$ \frac {-3560633}{196608}  $&$-1156 $&$ \frac {598880301}{16384} $\\ \hline
$0, 0, 2 $&$ \frac {203}{12288}  $&$\frac {57767}{12288}  $&$984 $&$ \frac {-16896729}{1024}$\\ \hline 
$1, 3, 0 $&$ \frac {1765}{4096}  $&$ \frac {1238171}{2048}  $&$-25008 $&$ \frac {-2314042407}{5120} $\\ \hline
$2, 1, 1 $&$ -  \frac {731}{12288} \sqrt{5} $&$ -  \frac {9299}{192} \sqrt{5} $&$ - 26112 \sqrt{5} $&$ -  \frac {993521361}{5120} \sqrt{5} $\\ \hline
$4, 2, 0 $&$ \frac {87389}{24576}  $&$ \frac {-3264407}{3072}  $&$53304 $&$ \frac {-14559909801}{10240} $\\ \hline
$5, 0, 1 $&$ \frac {11639}{24576} \sqrt{5} $&$ -  \frac {783017}{12288} \sqrt{5} $&$ - 30504\sqrt{5} $&$  -  \frac {418632555}{2048} \sqrt{5} $\\ \hline
$7, 1, 0 $&$ \frac {-15697}{49152}  $&$\frac {-15882179}{24576}  $&$105516 $&$ \frac {-1983545931}{4096} $\\ \hline
$10, 0, 0 $&$ \frac {11553}{163840} $&$ \frac {2496411}{81920}  $&$ \frac {15129}{5}  $&$ \frac {-5248288143}{40960} $\\ \hline
\end{tabular} \\
{Table 1: Coefficients of the functions $F_1,F_{2,0},F_{3,0,0,0}$ and $G_{11}^{-1}$.}
\end{center}
\end{document}